\documentclass[prl,a4paper,twocolumn,nofootinbib]{revtex4-1}
\usepackage[utf8x]{inputenc}
\usepackage{amssymb}
\usepackage{amsmath}
\usepackage{graphicx}
\usepackage{psfrag}
\usepackage{latexsym}
\usepackage{hyperref}

\newcommand{\eq}{\begin{equation}}
\newcommand{\eqx}{\end{equation}}
\newcommand{\eqn}{\begin{eqnarray}}
\newcommand{\eqnx}{\end{eqnarray}}

\newcommand{\f}[2]{\frac{#1}{#2}}

\begin{document}

\title{Hydrodynamic Gradient Expansion in Gauge Theory Plasmas}

\author{Michal P. Heller}\email{m.p.heller@uva.nl}
\altaffiliation[On leave from: ]{{\it National Centre for Nuclear Research,
  Ho{\.z}a 69, 00-681 Warsaw, Poland}}
\affiliation{\it Instituut voor Theoretische Fysica, Universiteit van Amsterdam, Science Park 904, 1090 GL Amsterdam, The Netherlands}

\author{Romuald A. Janik}\email{romuald@th.if.uj.edu.pl}

\author{Przemys{\l}aw Witaszczyk}\email{bofh@th.if.uj.edu.pl}

\affiliation{Institute of Physics,
Jagiellonian University, Reymonta 4, 30-059 Krak\'ow, Poland}

\begin{abstract}
We utilize the fluid-gravity duality to investigate the large order behavior of 
hydrodynamic gradient expansion of the dynamics of a gauge theory plasma system.
This corresponds to the inclusion
of dissipative terms and transport coefficients of very high order. 
Using the dual gravity description, we calculate numerically the form of the stress 
tensor for a boost-invariant flow in a hydrodynamic expansion up to terms 
with 240 derivatives.
We observe a factorial 
growth of gradient contributions at large orders, which indicates a zero radius of 
convergence of the hydrodynamic series. 
Furthermore, we identify the leading singularity in the Borel transform of the 
hydrodynamic energy density with the lowest nonhydrodynamic excitation corresponding to
a `nonhydrodynamic' quasinormal mode on the gravity side.
\end{abstract}

\maketitle

\noindent{\bf Introduction.} Hydrodynamics is an effective theory of fluids describing long-wavelength evolution of conserved quantities, among which are energy (local temperature) and momentum (local velocity). In the relativistic setting of uncharged hydrodynamics, which we consider here, one starts with the perfect fluid stress tensor and systematically corrects it by including all available gradient terms (dissipative corrections) with their relevance decreasing with the number of derivatives they carry. Each gradient term enters with an \emph{a priori} distinct function of temperature -- transport coefficient, e.g., the first order symmetric traceless contribution to the stress tensor is associated with the shear viscosity.

Transport coefficients are, in principle, derivable from the microscopic theory 
underlying the given fluid, being it kinetic theory for fluids formed out of weakly 
coupled quantum fields or the gauge-gravity duality for certain strongly coupled 
liquids. The exact values of transport coefficients are, of course, important for
the understanding of the dynamics of the particular fluid.
A famous example is the shear viscosity of quark-gluon plasma created in RHIC and LHC experiments, which needs to be sufficiently small in order to account for the observed spectra of particles (see \cite{HeinzSnellings} for a recent review).

Given the ubiquity and universality of the hydrodynamic description of many 
physical systems, it is interesting to investigate the character of the hydrodynamic gradient expansion
and its interrelation with nonhydrodynamic collective excitations or degrees of freedom. 

On a formal level, it is important to understand the large-order behavior of any 
approximate scheme in physics based on the existence of a small parameter such as hydrodynamic gradient expansion. Given the 
complexity of typical perturbative calculations at large orders, the question we are 
asking is challenging enough to deserve attention and interest. 
In fact, no methods were available so far to proceed effectively beyond the first few terms
in the hydrodynamic gradient expansion.

The issues of high order hydrodynamics are also interesting from a phenomenological point of view, in the context of ongoing relativistic heavy ion collision 
programs~\cite{HeinzSnellings}. Two examples considered recently in the literature are refining the criterium for the applicability of hydrodynamics by including terms with more than two derivatives \cite{Heller:2011ju,Heller:2012je} and trying to resum hydrodynamic expansion in order to account for situations in which individual gradient terms are not small enough to truncate the series at lowest orders \cite{Lublinsky:2007mm,Lublinsky:2009kv}.

In this Letter, we provide strong evidence that the radius of convergence of the hydrodynamic gradient expansion is zero. Furthermore, using the standard, in the context of divergent series, technique of Borel transform, we identify the leading singularity in Borel-transformed hydrodynamic stress tensor with the lowest lying nonhydrodynamic degree of freedom. As a byproduct, we uncover an intriguing structural
similarity of the phenomena arising in perturbative expansions in coupling
constant in quantum systems with temporal evolution governed by hydrodynamic
and nonhydrodynamic modes.

As a way of generating a hydrodynamic stress tensor at high orders of gradient expansion, we utilize the AdS/CFT correspondence \cite{Maldacena:1997re} and the fluid-gravity duality~\cite{Bhattacharyya:2008jc}. We obtain a gradient-expanded stress tensor for 
a particular solution of the equations of (all-order) relativistic hydrodynamics, 
generalizing
the so-called Bjorken flow~\cite{Bjorken:1982qr,Janik:2005zt} - a toy model of expanding quark-gluon plasma. The underlying microscopic theory is a strongly coupled ${\cal N}=4$ super Yang-Mills theory at large number of colors ($N_c$) and at strong coupling. The lowest lying nonhydrodynamic degree of freedom is then the metric quasinormal mode sharing the symmetries of the problem and having the lowest frequency.

\noindent{\bf Bjorken flow and holography.} Bjorken flow is a particular solution of the equations of relativistic hydrodynamics and describes matter expanding in one dimension with the additional assumption of boost-invariance in the longitudinal direction. This symmetry can be made manifest upon passing to curvilinear proper time $\tau$ - rapidity $y$ coordinates related to the lab frame time $x^{0}$ and position along the expansion axis $x^{1}$ via
\eq
\label{eq.tauy}
x^{0} = \tau \, \cosh{y} \quad \mathrm{and} \quad x^{1} = \tau \, \sinh{y}.
\eqx
In the case of (3+1)-dimensional conformal field theory plasma, the most general stress tensor obeying the symmetries of the problem in coordinates $(\tau, y, x^{1}, x^{2})$ reads
\eq
\label{eq.Tmunu}
T^{\mu}_{\,\, \nu} = \mathrm{diag} (-\epsilon, p_{L}, p_{T}, p_{T} )^{\mu}_{\,\, \nu},
\eqx
where the energy density $\epsilon$ is a function of proper time only and 
the longitudinal $p_{L}$ and transverse $p_{T}$ pressures are fully expressed in terms of the energy density \cite{Janik:2005zt}
\eq
p_{L} = - \epsilon - \tau \, \epsilon'  \quad \mathrm{and} \quad p_{T} = \epsilon + \frac{1}{2} \tau \, \epsilon'.
\eqx
Note that, in the proper time - rapidity coordinates \eqref{eq.tauy}, there is no momentum flow in the stress tensor \eqref{eq.Tmunu} and so the flow velocity is trivial and takes the form $u = \partial_{\tau}$. Hydrodynamic constituent relations lead, then, to gradient expanded energy density of the form
\eq
\label{eq.enden}
\epsilon = \frac{3}{8} N_{c}^2 \pi^2 \frac{1}{\tau^{4/3}} \left(\epsilon_{2} + \epsilon_{3} \frac{1}{\tau^{2/3}} + \epsilon_{4} \frac{1}{\tau^{4/3}} + \ldots\right),
\eqx
where the choice of $\epsilon_{2}$ sets an overall energy scale, in particular for the quasinormal frequencies (\ref{eq.qnmfrequency}) and \ref{eq.closestpole}). The prefactor was chosen to match the ${\cal N} = 4$ super Yang-Mills theory at large-$N_{c}$ and strong coupling. In the following, we choose the units by setting $\epsilon_{2} = \pi^{-4}$.

Large-$\tau$ expansion of the energy density in powers of $\tau^{-2/3}$, as in \eqref{eq.enden}, is equivalent to the hydrodynamic gradient expansion and arises from expressing gradients of velocity ($\nabla_{\mu} u_{\nu} \sim \tau^{-1}$) in units of the effective temperature ($T \sim \epsilon^{1/4} \sim \tau^{-1/3}$). The value of the coefficient $\epsilon_{3}$ is related to the shear viscosity $\eta$, whereas $\epsilon_{4}$ is a sum of two transport coefficients: relaxation time $\tau_{\Pi}$ and the so-called $\lambda_{1}$ \cite{Baier:2007ix}. Higher order contributions to the energy density are expected to be linear combinations of so far unidentified transport coefficients. Note also that the expansion \eqref{eq.enden} is sensitive to both linear and nonlinear gradient terms.

As explained in \cite{Heller:2008mb,Kinoshita:2008dq} (see also Supplemental material), higher order contributions to the energy density \eqref{eq.enden} can be obtained by solving Einstein's equations with a negative cosmological constant for the metric ansatz of the form
\eq
\label{eq.metricansatz}
ds^{2} = 2 d\tau dr - A d\tau^2 + \Sigma^{2} e^{-2 B} d y^2 + \Sigma^2 e^{B} (dx_{1}^2 + dx_{2}^2),
\eqx
where the warp factors $A$, $\Sigma$ and $B$ are functions of $r$ and $\tau$ constructed in the gradient expansion as required by the fluid-gravity duality. At leading order, the warp factors are that of a locally boosted black brane and this solution gets systematically corrected in $\tau^{-2/3}$ expansion, as is the case with the energy density in the dual field theory \eqref{eq.enden}.

The background expanded in $\tau^{-2/3}$ around a locally boosted black brane is slowly evolving and captures only hydrodynamic degrees of freedom. One can, in addition, consider the incorporation of nonhydrodynamic (fast evolving) degrees of freedom by linearizing Einstein's equations on top of the hydrodynamic solution, i.e.
$B= B_{\mathrm{hydro}} + \delta B$, and similarly for $A$ and $\Sigma$, and looking for $\delta B$ corresponding to (at very large time) the exponentially decaying contribution to the stress tensor depending only on $\tau$. For the static background analogous calculation would lead to the spectrum of nonhydrodynamic quasinormal modes carrying zero momentum, which is known to be the same as the spectrum of zero momentum quasinormal modes for the massless scalar field \cite{Kovtun:2005ev}.

In the leading order of the gradient expansion, the resulting modes, on the gravity side, indeed essentially reduce to the scalar quasinormal modes but obtain an additional
factor of $\f{3}{2}$ and are damped exponentially in $\tau^{\f{2}{3}}$ 
\cite{Janik:2006gp}. Upon including viscous correction, the modes obtain a 
further nontrivial powerlike preexponential factor

\vspace{-0.5cm}

\eq
\label{eq.qnmcontrib}
\delta \epsilon \sim \tau^{\alpha_{qnm}} \exp{(- i \, \f{3}{2} \, \omega_{qnm} \, \tau^{2/3})}.
\eqx

\vspace{-0.1cm}

Explicit gravity calculation for the lowest mode yield

\vspace{-0.6cm}

\eq
\label{eq.qnmfrequency}
\omega_{qnm} = 3.1195 -2.7467, \,\, \alpha_{qnm} = -1.5422 + 0.5199 \, i.
\eqx

\vspace{-0.1cm}

The frequency $\omega_{qnm}$ agrees with the frequency of the lowest nonhydrodynamic scalar quasinormal mode and was calculated before in \cite{Janik:2006gp}, whereas the prediction of $\alpha_{qnm}$ is a new result specific to the dissipative modifications of the expanding black hole geometry (see the Supplemental Material for further details).
In the following, we will be able to reproduce numerically \eqref{eq.qnmfrequency} 
just from the large order behavior of the hydrodynamic series.

\noindent{\bf Large order behavior of hydrodynamic energy density.} 
Numerical implementation of the methods outlined in \cite{Heller:2008mb,Kinoshita:2008dq} allow for efficient calculation of hydrodynamic series given by \eqref{eq.enden}, up to a very large order, since one is effectively solving a set of linear ODE's (coming from Einstein's equations) at each order. 
Using spectral methods we iteratively solved these equations in the large time expansion reconstructing the energy density up to the order 240, i.e. up to the term $\epsilon_{242}$ in \eqref{eq.enden}. To the best of our knowledge this is the first approach allowing us to access information about the large order behavior of the hydrodynamic series in any physical system or model.

As a way of monitoring the accuracy of our procedures we compared normalized values of evaluated Einstein's equations at each order of the $\tau^{-2/3}$ expansion to the ratio of coefficients of gradient-expanded energy density to gradient expanded warp factors. This ensures that our results for the energy density are reliable. 
We also verified that we reproduced the known analytic results for the energy density at low ($\leq 3$) orders.

\begin{figure}
\includegraphics[width=7.3cm]{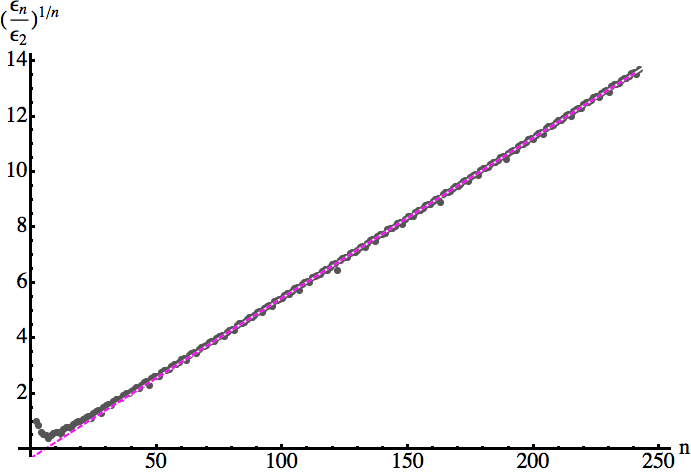}
\caption{Behavior of the coefficients of hydrodynamic series for the energy density as a function of the order. At large enough order the coefficients start exhibiting factorial growth. The radius of convergence of the Borel transformed series is estimated to be $6.37$, 
in rough agreement with \eqref{eq.closestpole}: $2/3\times6.37 = 4.25$.}
\label{fig:asbhvr}
\end{figure} 

As anticipated in the introduction, the coefficients of the gradient expanded energy density \eqref{eq.enden} that we obtained, indicate that the hydrodynamic gradient expansion, as seen in the example of boost-invariant flow, has zero radius of convergence. This is clearly visible on Fig.~\ref{fig:asbhvr}, which shows the factorial growth with order of the coefficients of the hydrodynamic series for the energy density \eqref{eq.enden}. Numerical values of coefficients $\epsilon_{n}$, as in Eq.~\eqref{eq.enden}, can be found in the file ``eps.m'' included in the submission (see also Supplemental Material).

\noindent{\bf Singularities of the Borel transform and quasinormal modes.} A standard way of dealing with divergent series, including perturbative quantum field theories \cite{Suslov:2005zi}, is performing a Borel transform of the original power 
series (here a series in $u \equiv \tau^{-\f{2}{3}}$), 
$\tilde{\epsilon}_{n} = \epsilon_{n} / \, n!$,
and subsequently the Borel resummation
\eq
\label{eq.borelsum}
\epsilon_{resum}(\tau)=\int_{0}^{\infty}\tilde{\epsilon}(\zeta \tau)\, 
e^{-\zeta}\, d\zeta,
\eqx
where the integral is taken over the real positive axis.
A~key issue in performing Borel resummation is the identification of the 
singularities of the Borel transform and its analytical continuation
into some neighborhood of the positive real axis. The singularities
of the Borel transform are interesting for their own sake as, typically, 
they have a definite physical interpretation (like instanton contributions
in quantum field theories). We will find that this will also be the case
in our context.

Since the Borel transform has typically only a finite radius of
convergence, we use the standard technique of Pad\'e approximants to
provide an analytic continuation. Such an approach was used with success in the context of resumming perturbative series in quantum field theories in \cite{Ellis:1995jv} and, in principle, can be refined by including the information about the behavior of the series at large values of the parameter (typically strong coupling behavior for perturbative expansion in quantum field theories).

In our case, this regime corresponds to small times (high order dissipative terms). Although it was established that, when expanded around $\tau = 0$, the energy density contains only even powers of proper time \cite{Beuf:2009cx}, this information turns out to be hard to implement in an unambiguous way and, in the following, we adopted the simplest analytic continuation.

\begin{figure}
\includegraphics[width=8cm]{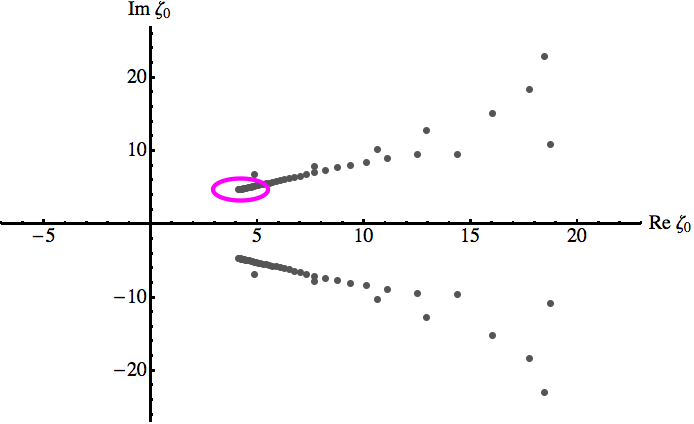}
\caption{Real and imaginary parts of poles $\zeta_{0}$ of the symmetric Pad\'e approximant of the Borel transform of the energy density \eqref{eq.enden}. From the plot we removed numerically spurious poles. The pole closest to the origin governs the convergence radius of the Borel transformed series and gives rise to the lowest quasinormal frequency. The poles from the encircled region (magenta) lead to the powerlike preexponential factor in \eqref{eq.qnmcontrib}.}
\label{fig:poles}
\end{figure} 

Figure \ref{fig:poles} shows the position of poles of the symmetric Pad\'e approximant of the Borel transformed energy density. Let us note that some care must be taken as the Pad\'e approximant exhibits apparent poles on the negative real axis which, however,
cancel almost perfectly (up to $10^{-100}$ accuracy) with zeroes of the numerator. 
As our knowledge about the series is limited to a finite number of terms (first 241), we should only trust the structure in the close vicinity of the origin. 
Note, however, that the lack of poles on the positive real axis seems to indicate Borel summability, pointing towards the possible existence of Borel-resummed 
all-order hydrodynamics.

The poles approximate some complicated structure of branch cuts. The pole nearest the origin, from which a major branch cut starts, sets the radius of convergence of the Borel transform of hydrodynamic series. Its numerical value (multiplied by the factor $3/2 i$) reads 
\eq
\label{eq.closestpole}
\omega_{\mathrm{Borel}} =3.1193 - 2.7471 \,i \quad (|\omega_{\mathrm{Borel}}| = 4.1565)
\eqx
and is consistent with the coefficient obtained from fitting factorial-like behavior of the original series.

Deforming the contour from the real positive axis to the one encircling some of the zeros of the denominator of the Pad\'e approximated Borel transform, through residues, leads to an ambiguity in resummation. 
It is tempting to speculate that such an ambiguity can be understood as 
the appearance of new, nonhydrodynamic, degrees of freedom in the system. Indeed, performing contour integration around the cut originating from $(3/2 i) \times \omega_{\mathrm{Borel}}$ leads to the following contribution to the energy density at large proper time
\eq
\label{eq.deltaeps}
\delta \epsilon \sim \tau^{\alpha_{\mathrm{Borel}}} \exp{(- i \, \frac{3}{2} \, \omega_{\mathrm{Borel}} \, \tau^{2/3})},
\eqx
where
\eq
\label{eq.alphaandomegafromPade}
\alpha_{\mathrm{Borel}} = -1.5426+0.5192 \,i.
\eqx
The contribution \eqref{eq.deltaeps} together with \eqref{eq.closestpole} and \eqref{eq.alphaandomegafromPade} matches, up to numerical accuracy, the gravitational prediction for the behavior of the lowest metric quasinormal mode given by \eqref{eq.qnmcontrib} and \eqref{eq.qnmfrequency}, but is derived entirely within the hydrodynamic gradient expansion.

\noindent{\bf Summary and conclusions.} We used the fluid-gravity duality to numerically obtain the form of hydrodynamic stress tensor of the Bjorken flow up to $240^{\mathrm{th}}$ order in gradients. This corresponds to the inclusion of dissipative terms of a very high order. We discovered that the corrections grow factorially with the order. This provides strong indication that hydrodynamic expansion is an example of asymptotic series; i.e., it has a zero radius of convergence.

Upon a simple analytic continuation of Borel transformed series we found that it is the frequency of the lowest nonhydrodynamic quasinormal mode which sets the radius of convergence of the Borel transform. Subsequently, from the hydrodynamic expression alone we extracted the modification of this contribution due to the shear viscosity and matched with the gravity calculation. It is fascinating to observe that
nonhydrodynamic collective excitations, which are not describable within
the framework of hydrodynamics, nevertheless leave their imprint in the
behavior of high order transport coefficients. This is reminiscent of the
relation of instanton contributions with the divergence of the perturbative
series in quantum field theory.

Note that our observation about the asymptotic character of the hydrodynamic series is made upon evaluating the stress tensor of a particular type of hydrodynamic
flow. The high degree of symmetry of that flow -- boost-invariance -- means
that at each order in the large proper time expansion, we are indeed studying 
new dissipative terms in the hydrodynamic stress tensor with new transport
coefficients. The dependence on initial conditions for this hydrodynamic
flow reduces just to a trivial overall rescaling.

We speculate that the most likely source of the factorial growth of the coefficients of the hydrodynamic series for the energy density is the fast growth of the number of gradient terms contributing to the hydrodynamic stress tensor at each order,
in analogy with the factorial growth of the number of Feynman graphs at large order of perturbative calculations.

Regarding future directions, a thought-provoking phenomenological spin off of our Letter is the possibility of resumming hydrodynamic description and extending the hydrodynamic stress tensor past the regime in which subsequent low order terms give comparable contributions. This might lead to refined criterium for the applicability of hydrodynamics used in \cite{Heller:2011ju,Heller:2012je} and goes very much in the spirit of \cite{Lublinsky:2007mm,Lublinsky:2009kv}. 

\noindent{\bf Acknowledgments.} MPH acknowledges support from the Netherlands Organization for Scientific Research under the Veni scheme and would like to thank Nordita for hospitality. 
This work was supported by NCN grant 2012/06/A/ST2/00396 (RJ) and IoP JU grant K/DSC/000705 (PW). PW would like to thank the Lorentz Center and Nordita for hospitality. 
We used M. Headrick’s \href{http://people.brandeis.edu/~headrick/Mathematica/index.html}{\tt diffgeo.m} package for symbolic GR. Finally we would like to thank M. Baggio, J. de Boer, K. Landsteiner, C. Nunez and D. Teaney for interesting discussions and suggestions.

\section*{Supplemental material}

\noindent{\bf Gravity solution and dual energy density.} 
The metric ansatz \eqref{eq.metricansatz} has a residual diffeomorphism freedom $r \rightarrow r + f(\tau)$ \cite{Kinoshita:2008dq}. The near-boundary expansion of \eqref{eq.metricansatz} contains information about the expectation of the dual stress tensor. For $f(\tau) = 0$ the metric close to the boundary take the form
\eqn
A &=& r^2 + a_{4}(\tau) r^{-2} + \ldots \\
B &=& \frac{2}{3} \log{\frac{r}{1 + r t}} - \frac{1}{2} [a_{4}(\tau) + \frac{3}{4} \tau a_{4}'(\tau)] \, r^{-4} + \ldots \nonumber
\eqnx
where $a_{4}$ is related to the (local) energy density of the plasma
\eq
\epsilon(\tau) = - \frac{3}{8 \pi^{2}} N_{c}^{2} \, a_{4}(\tau).
\eqx
We look for the solution in the form dictated by the fluid-gravity duality, i.e.
\eqn
\label{eq.gradientexpandedwarps}
A &=& r^2 \sum_{j = 0} \frac{1}{\tau^{2/3 j}} A_{j}(r \, \tau^{1/3}), \nonumber \\
B &=& \frac{2}{3} \log{\frac{r}{1 + r t}} + \sum_{j = 0} \frac{1}{\tau^{2/3 j}} B_{j}(r \, \tau^{1/3}) ,\nonumber \\
\Sigma &=& r^{2/3} (1 + r t)^{1/3} \sum_{j = 0} \frac{1}{\tau^{2/3 j}} \Sigma_{j}(r \, \tau^{1/3}),
\eqnx
where the starting point is the boosted black brane metric
\eq
\label{eq.0order}
A_{0}(v) = 1 - \frac{1}{v^{4}}, \quad B_{0}(v) = 0 \quad \mathrm{and} \quad \Sigma_{0}(v) = 1.
\eqx

\noindent{\bf The choice of units and provided data for the energy density.} In \eqref{eq.0order} we chose the units by setting its event horizon to lie at $r = 1$ and this choice leads to $\epsilon_{2} = \pi^{-4}$ in the main body of the paper. The file ``eps.m'', included in the submission, contains coefficients of the energy density in the late time expansion written in {\tt Mathematica} format as $\{j,\epsilon_{j}\}$. 

In order to restore general units, i.e. bring the energy density to the form
\eq
\epsilon (\tau) = \frac{3}{8}  \pi^{2} N_{c}^{2} \frac{\Lambda^{4}}{(\Lambda \tau )^{4/3}}\left\{c_{2} + c_{3} \frac{1}{(\Lambda \tau )^{2/3}} + \ldots  \right\},
\eqx
where $\Lambda$ is an overall scale and $c_{2} = 1$, one needs to multiply each $\epsilon_{j>2}$ that we provide by $\pi^{6-j}$
\eq
\epsilon_{j} = \pi^{j-6} \, c_{j}.
\eqx

\noindent{\bf Numerical implementation.} At each order order of $\frac{1}{\tau^{2/3}}$ expansion we are solving a set of coupled linear ordinary differential equations for 3 functions. There are 5 different equations and we explicitly solve 3 of them and use the remaining 2 to provide boundary conditions for the functions at $v = 1$. In the numerical implementation we fixed the residual diffeomorphism freedom by demanding that $A_{j}(1) = 0$ for all $j$. The bulk regularity requires that the mode diverging logarithmically in the vicinity of the horizon vanishes. The flatness of the boundary metric provides the remaining boundary conditions for the metric at each order. See \cite{Kinoshita:2008dq} for detailed discussion of the late time gravity solution to the Bjorken flow in the Eddington-Finkelstein coordinates.

At each order we generated the equations symbolically. The kernels for all equations take particularly simple form and the whole complication arises from source terms which significantly grow in length with order. We solved these equations in {\tt Mathematica} via matrix inversion using spectral discretization in $v$ direction [using $z = 1/v$ running from $0$ (the boundary) to $1$ (the horizon)]. The energy density at each order is obtained by taking the fourth derivative of $A_{j}(1/z)$ at $z = 0$.

Going to $240^{\mathrm{th}}$ order required a grid of $250$ points with extended precision (keeping the first 450 digits) and took about 4 weeks on a desktop computer with 3.4~GHz processor and 16 GB of memory.

\noindent{\bf QNMs on top of the expanding plasma.} Small perturbations on top of the expanding plasma are captured by the linearized Einstein's equations
\eq
B = B_{\mathrm{hydro}} + \delta B,
\eqx
with analogous expressions for $A$ and $\Sigma$. The fluid-gravity duality dictates that at late time the perturbations coincide with corresponding perturbations of the static black brane with the temperature replaced by the local temperature of the flow \cite{Janik:2006gp}. Within the gradient expansion one thus expects that
\eqn
\label{eq.qnmpert}
\delta B = && \exp{[- i \int d \tau (\omega_{0} \tau^{-1/3} + \omega_{1} \tau^{-1} + \ldots]} \times \nonumber \\
&& \left\{\delta B_{0} (r \, \tau^{1/3}) + \frac{1}{\tau^{2/3}} \delta B_{1} (r \, \tau^{1/3}) + \ldots \right\}
\eqnx
and similar expressions for $\delta A$ and $\delta \Sigma$.

Upon linearizing equations on top of \eqref{eq.gradientexpandedwarps} and keeping terms up to the first order in gradients one discovers that $\delta B$ given by \eqref{eq.qnmpert} obeys the same equation as the massless scalar field. Subleading correction takes into account viscous effects present in the expanding plasma described by hydrodynamics.

The overall normalization does not matter, which allows to set $\delta B_{0}(1) = 1$. This choice fixes both the normalization and chooses the ingoing boundary condition at the horizon. Imposing Dirichlet boundary condition at $v = \infty$ leads to the discrete spectrum of frequencies the lowest one giving
\eq
\omega_{0} = \omega_{qnm} = 3.1195 - 2.7467 i,
\eqx
which is the value quoted in the main body of the article.

Repeating the analysis in the first order allows to obtain
\eq
\omega_{1} = -0.5199 + 0.4578 i.
\eqx
It is perhaps worth stressing that the value of $\delta B_{1}(v)$ at $v = 1$ turns out not to matter for obtaining $\omega_{1}$, as the associated solution vanishes at the boundary.

Finally, in order to obtain the correction to the energy density $\delta \epsilon (\tau)$, one needs to take into account that the asymptotic behavior of $\delta B_{1}$ is indirectly related to the energy density $\delta \epsilon(\tau)$ through
\eq
\label{eq.dBnearbdry}
\delta B \Big |_{r \rightarrow \infty} \sim \left\{\delta \epsilon(\tau) + \frac{3}{4} \tau \delta \epsilon'(\tau)\right\} r^{-4}.
\eqx
Solving this relation in the late time expansion is straightforward and leads to
\eq
\label{eq.alphaqnm}
\alpha_{qnm} = - i \omega_{1} - \frac{4}{3} - \frac{2}{3} = - 1.5422 + 0.5199 i,
\eqx
as quoted in the main body of the article. The factor of $-4/3$ in \eqref{eq.alphaqnm} follows from passing from the scaling variable $v = r \,\tau^{1/3}$ to the original radial variable $r$ and the factor of $-2/3$ follows from solving \eqref{eq.dBnearbdry} for $\delta \epsilon(t)$ in the late time expansion.

The match between the gravity and hydrodynamic calculation of the lowest quasinormal mode (including the viscous correction) provides a highly nontrivial check of the validity of both calculations.


\begin{thebibliography}{99}

\bibitem{HeinzSnellings} 
  U.~W. Heinz and R.~Snellings,
  arXiv:1301.2826 [nucl-th].

\bibitem{Heller:2011ju} 
  M.~P.~Heller, R.~A.~Janik and P.~Witaszczyk,
  Phys.\ Rev.\ Lett.\  {\bf 108}, 201602 (2012)
  [arXiv:1103.3452 [hep-th]].
  
\bibitem{Heller:2012je} 
  M.~P.~Heller, R.~A.~Janik and P.~Witaszczyk,
  Phys.\ Rev.\ D {\bf 85}, 126002 (2012)
  [arXiv:1203.0755 [hep-th]].
  
\bibitem{Lublinsky:2007mm}
  M.~Lublinsky and E.~Shuryak,
  Phys.\ Rev.\  C {\bf 76}, 021901 (2007)
  [arXiv:0704.1647 [hep-ph]].

\bibitem{Lublinsky:2009kv} 
  M.~Lublinsky and E.~Shuryak,
  Phys.\ Rev.\ D {\bf 80}, 065026 (2009)
  [arXiv:0905.4069 [hep-ph]].
 
\bibitem{Maldacena:1997re}
  J.~M.~Maldacena,
  Adv.\ Theor.\ Math.\ Phys.\  {\bf 2}, 231 (1998)
  [Int.\ J.\ Theor.\ Phys.\  {\bf 38}, 1113 (1999)]
  [arXiv:hep-th/9711200].

\bibitem{Bhattacharyya:2008jc}
  S.~Bhattacharyya, V.~E.~Hubeny, S.~Minwalla and M.~Rangamani,
  JHEP {\bf 0802}, 045 (2008)
  [arXiv:0712.2456 [hep-th]].

\bibitem{Bjorken:1982qr}  J.~D.~Bjorken,
  Phys.\ Rev.\ D {\bf 27}, 140 (1983).

\bibitem{Janik:2005zt}
  R.~A.~Janik and R.~B.~Peschanski,
  Phys.\ Rev.\  D {\bf 73}, 045013 (2006)
  [arXiv:hep-th/0512162].

\bibitem{Baier:2007ix} 
  R.~Baier, P.~Romatschke, D.~T.~Son, A.~O.~Starinets and M.~A.~Stephanov,
  JHEP {\bf 0804}, 100 (2008)
  [arXiv:0712.2451 [hep-th]].
 
\bibitem{Heller:2008mb} 
  M.~P.~Heller, P.~Surowka, R.~Loganayagam, M.~Spalinski and S.~E.~Vazquez,
  Phys.\ Rev.\ Lett.\  {\bf 102}, 041601 (2009)
  [arXiv:0805.3774 [hep-th]].

\bibitem{Kinoshita:2008dq} 
  S.~Kinoshita, S.~Mukohyama, S.~Nakamura and K.~-y.~Oda,
  Prog.\ Theor.\ Phys.\  {\bf 121}, 121 (2009)
  [arXiv:0807.3797 [hep-th]].

\bibitem{Kovtun:2005ev} 
  P.~K.~Kovtun and A.~O.~Starinets,
  Phys.\ Rev.\ D {\bf 72}, 086009 (2005)
  [hep-th/0506184].

\bibitem{Janik:2006gp} 
  R.~A.~Janik and R.~B.~Peschanski,
  Phys.\ Rev.\ D {\bf 74}, 046007 (2006)
  [hep-th/0606149].
  
\bibitem{Suslov:2005zi} 
  I.~M.~Suslov,
  Zh.\ Eksp.\ Teor.\ Fiz.\  {\bf 127}, 1350 (2005)
  [J.\ Exp.\ Theor.\ Phys.\  {\bf 100}, 1188 (2005)]
  [hep-ph/0510142].


\bibitem{Ellis:1995jv} 
  J.~R.~Ellis, E.~Gardi, M.~Karliner and M.~A.~Samuel,
  Phys.\ Lett.\ B {\bf 366}, 268 (1996)
  [hep-ph/9509312].

\bibitem{Beuf:2009cx} 
  G.~Beuf, M.~P.~Heller, R.~A.~Janik and R.~Peschanski,
  JHEP {\bf 0910}, 043 (2009)
  [arXiv:0906.4423 [hep-th]].


\end{thebibliography}
\end{document}